\documentclass[11pt,twoside]{article}
\usepackage{cozumel2005}
\usepackage{epsf}
\usepackage{psfig}
\usepackage{graphicx}
\usepackage{lscape}
\begin{document}
\title{The SFH of the LMC: The CMD approach vs. integrated colors and spectra}    
\author{Thomas Lilly and Uta Fritze-v. Alvensleben}   
\affil{Institut f\"ur Astrophysik, Universit\"at G\"ottingen, Germany}    

\begin{abstract} 
We present results of a study aiming at shedding light on the specific advantages and limitations of methods to
derive star formation histories (SFH) in galaxies using resolved stellar populations and using integrated light,
respectively. For this purpose, we analyse the integrated-light spectrum of a field in the LMC bar, for which
highly resolved HST images are available as well. To be compared with the SFH derived from the color magnitude
diagram (CMD) of this field (Smecker-Hane et al. 02), we have performed a set of simulations of galaxies with
systematically varying SFHs, but constant metallicity (Z=0.008). We investigate to which extent different SF
scenarios can be discriminated on the basis of their photometric and spectral properties, respectively, and
determine in how far the detailed SFH obtained by the CMD approach can be reproduced by results based upon
integrated properties. Comparing both methods for this nearby field we want to learn about the specific character
of both methods and understand to what precision SFHs can be determined for distant galaxies only observable in
integrated light.\\
All simulations are performed using our evolutionary synthesis code GALEV.
\end{abstract}




\section{Introduction}
The study presented here is part of our work within a larger collaborative project with the aim to confront
different methods to derive SFHs from integrated light against each other and against the color magnitude diagram
(CMD) approach.

Test object for this ongoing project is a field in the bar of the Large Magellanic Cloud (LMC), for which both an
integrated-light spectrum (obtained with the 3.6m ESO telescope, LaSilla) and data on its resolved stellar
population (obtained with the \emph{Hubble Space Telescope}) are available. That way, the results of the different
groups analysing the spectrum cannot only be compared with each other but can also be compared with the SFH
obtained by an analysis of the CMD of the same field.
A short description of the project can be found in Alloin et al. (2002); an analysis of the CMD for this field is
presented by Smecker-Hane et al. (2002).

In this contribution, we present parts of our analysis of the integrated-light spectrum of the LMC bar field.
Applying the results of our preparatory work presented in Lilly \& Fritze -- v. Alvensleben (2005, these
proceedings), and using our evolutionary synthesis code GALEV, we have performed a set of simulations of galaxies
with systematically varying SFHs, but constant metallicity (Z=0.008). By confrontation of the evolution of the
colors and spectra resulting from the various simulations we then investigate to which extent different SF
scenarios can be discriminated on the basis of their photometric and spectral properties, respectively, and in how
far the detailed SFH obtained by the CMD approach can be reproduced by integrated properties. To keep the study
focused, we keep other parameters, like the initial mass function or the metallicity, constant.

A short overview of our evolutionary synthesis code GALEV and its input physics can be found in Lilly \&
Fritze -- v. Alvensleben (2005, these proceedings).

\section{The LMC bar field and its spectrum}
\begin{figure}[t]
   \plottwo{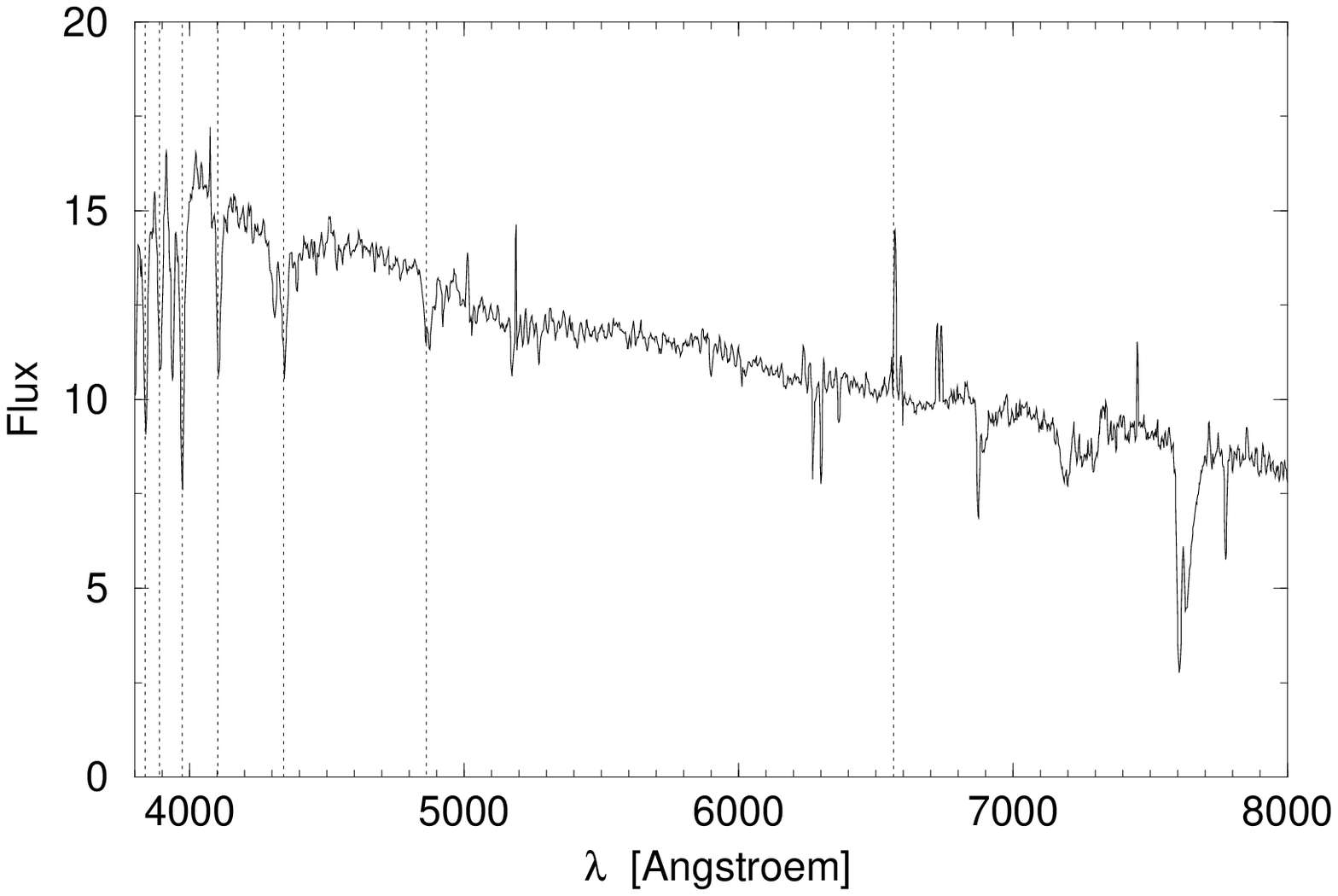}{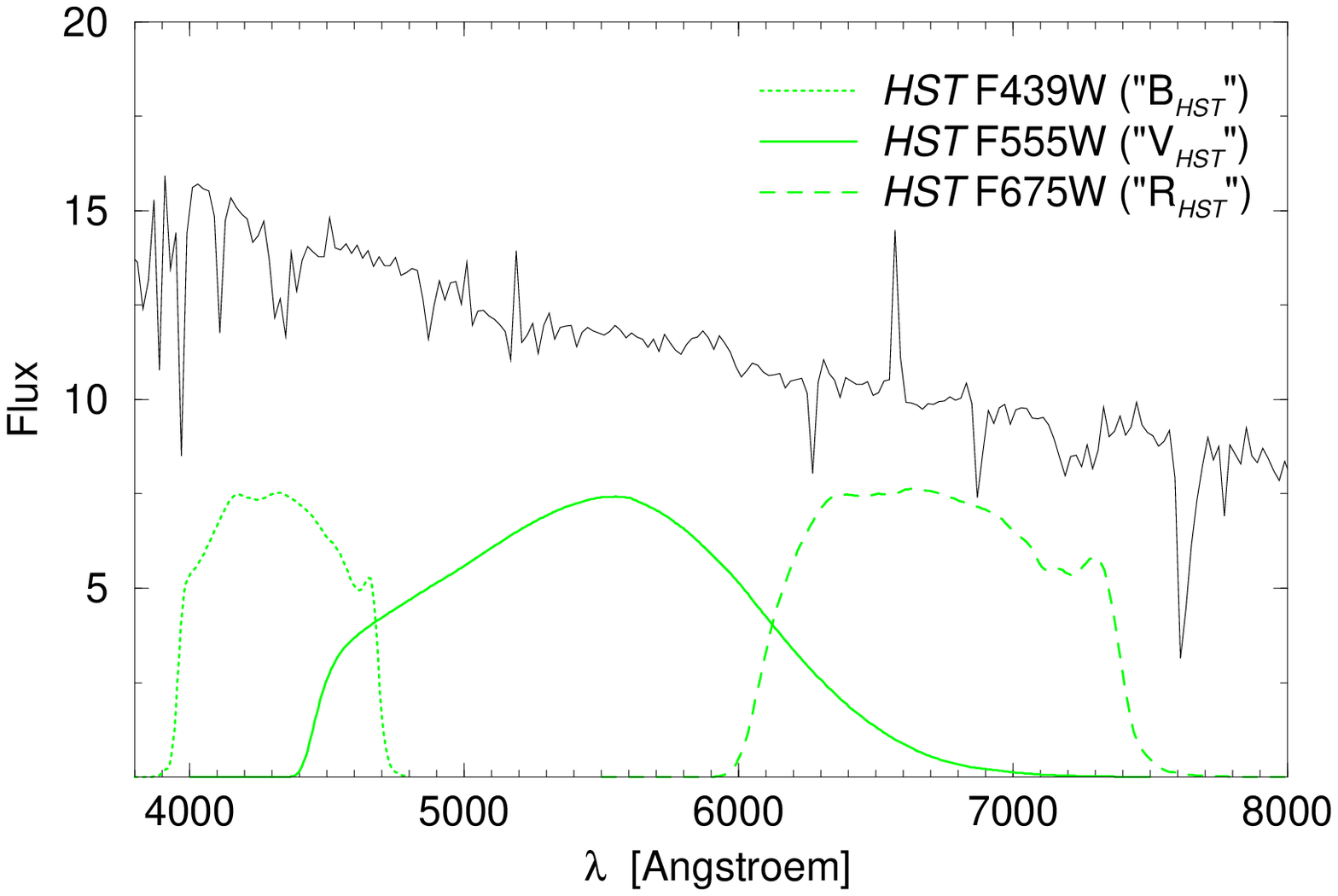}
   \caption{Integrated-light spectrum of the LMC bar field (dereddened). {\itshape Left:\/} Original spectrum.
   {\itshape Right:\/} Spectrum with lowered resolution to be compared with model spectra, and the three filters used
   for the analysis.}
\label{abb.lmcspec}
\end{figure}
The integrated-light spectrum of the LMC bar field (FoV: 2.5'$\times$5') was obtained at the ESO 3.6m telescope on
La Silla in Dezember 2000 by E. Pompei and D. Alloin. The 2000 coordinates of the field are: $\alpha$ = 05:23:17
and $\delta$ = --69:45:42; for observational details cf. Alloin et al. 2002.

Figure \ref{abb.lmcspec} (left) shows the original spectrum (dereddened with A$_V$ = 0.249), Figure
\ref{abb.lmcspec} (right) the same spectrum but with lowered resolution to be compared with our model spectra.
For better orientation, Balmer lines $H_\alpha$ to $H_\eta$ are marked by vertical lines in the left panel.

\begin{table}
   \caption{Colors derived from the integrated-light spectrum of the LMC bar field, obtained by folding the
   spectrum with the respective filter functions.}
   \smallskip
   \begin{center}
   {\small
   \begin{tabular}{c c c}
         \tableline
         \noalign{\smallskip}
         (B--V)$_{HST}$ & (V--R)$_{HST}$ & (B--R)$_{HST}$\\
         \noalign{\smallskip}
         \tableline
         \noalign{\smallskip}
         0.44mag & 0.53mag & 0.97mag\\
         \noalign{\smallskip}
         \tableline
   \end{tabular}
   }
   \end{center}
   \label{tab.colors}
\end{table}

\noindent Unfortunately, the model spectra we use have a resolution too low to successfully analyse spectral
features; therefore, our analysis mainly depends on broad band colors obtained from the spectrum, and the shape of
the spectral energy distribution.

The usable wavelength range of the spectrum allows for folding with three filters (filter functions shown in Fig.
\ref{abb.lmcspec}, right panel):
\emph{HST} WFPC2 F439W, F555W, and F675W (in the following referred to as B$_{HST}$, V$_{HST}$, R$_{HST}$);
Table \ref{tab.colors} gives the resulting colors obtained by folding the spectrum with the respective filter
functions.

\section{A simple 3-phase SFH}
\begin{figure}[!t]
   \begin{center}
   \includegraphics[width=0.47\linewidth]{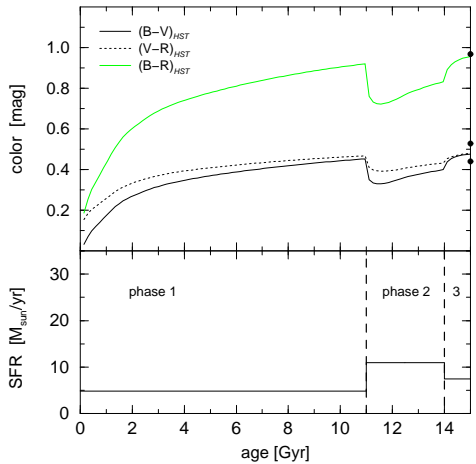}
   \caption{Photometric evolution of a simple 3-phase scenario in terms of (B--V)$_{HST}$, (V--R)$_{HST}$, and
   (B--R)$_{HST}$ with the corresponding SFH; the observed LMC bar field colors (cf. Table \ref{tab.colors}) are
   marked with black dots at 15 Gyr galaxy age.}
   \label{abb.lmc22}
   \end{center}
\end{figure}
\begin{figure}[t]
   \begin{center}
   \includegraphics[width=0.52\linewidth]{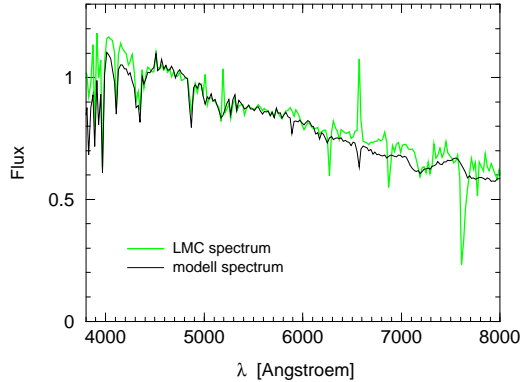}
   \caption{Model spectrum of the 3-phase scenario (cf. Fig. \ref{abb.lmc22}) at a galaxy age of 15 Gyr against
   the observed LMC spectrum. Both spectra are normalized at 4810\AA\ (arbitrary value).}
   \label{abb.lmc22spec}
   \end{center}
\end{figure}
From our preparatory work presented in Lilly \& Fritze -- v. Alvensleben (2005, these proceedings), we have learned
that, using integrated light only, variations in the SFH of a galaxy can be traced for only about 1, at the
utmost 4 Gyrs of lookback time.
This means, the relative distribution of SF \emph{within} early epochs ($\geq$ 4 Gyr ago) of galaxy evolution is
almost irrelevant, and that of the nearer history (4 Gyr $\geq$ lookback time $\geq$ 1 Gyr) only of weak relevance
for the observed colors; on the other hand, the relative distribution of SF \emph{between} these 3 phases -- and
not only the SF within the last Gyr -- \emph{is} important.
Therefore, it should be possible to fit any given set of colors (or, as we have also learned at the study,
even Lick indices) using an evolutionary history of only 3 different phases of SF, if the phases are chosen
appropriately:

Figures \ref{abb.lmc22} and \ref{abb.lmc22spec} show that a very simple ``3-phase scenario'' with phase 1 ranging
from 0 to 11 Gyr galaxy age, phase 2 from 11 to 14 Gyr galaxy age, phase 3 from 14 to 15 Gyr galaxy age, can
indeed fit both the observed colors (they are marked with black dots at the right edge of Fig. \ref{abb.lmc22},
top panel) and the observed spectrum (cf. Fig. \ref{abb.lmc22spec}) of the LMC bar field reasonably well;
differences still visible are most likely due to our disregarding of any chemical enrichment history and the fact
that we assume a constant SFR within phase 3.\\

\begin{figure}[t]
   \begin{center}
   \includegraphics[height=0.29\linewidth]{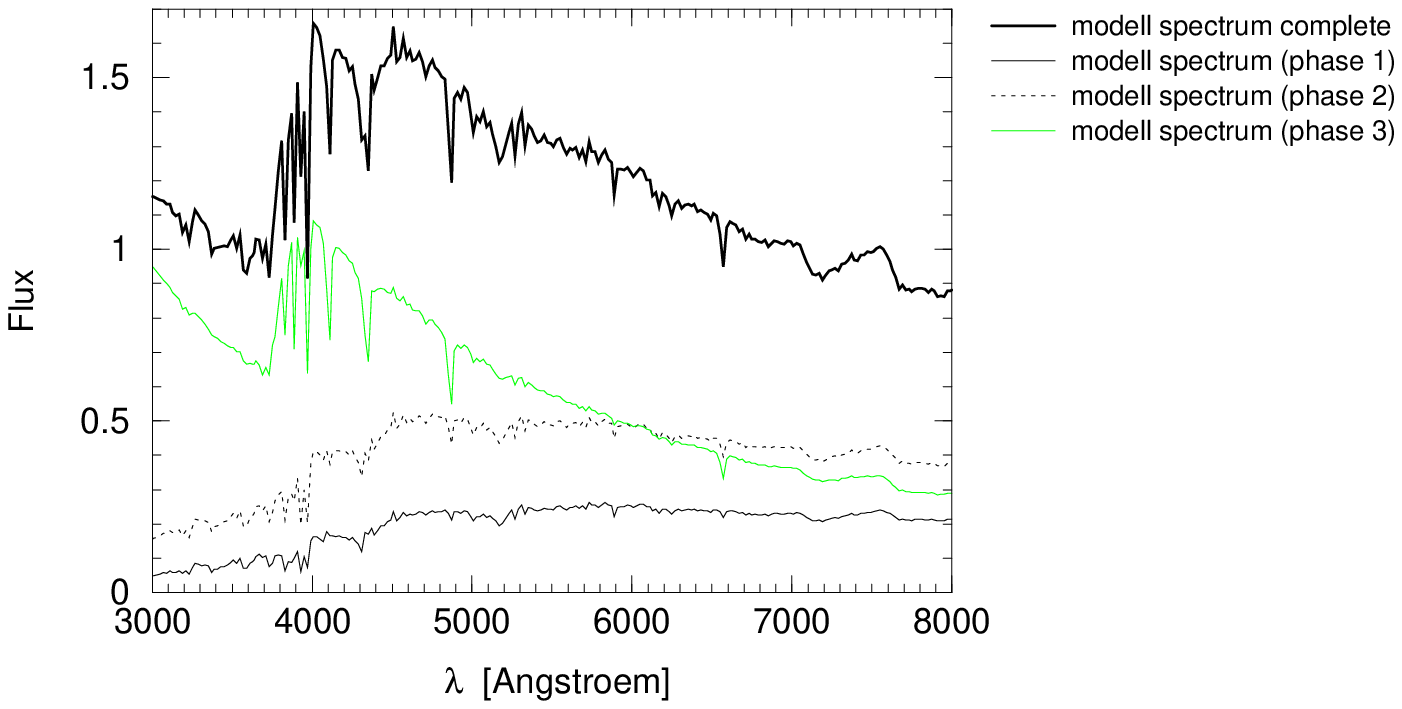}
   \includegraphics[height=0.29\linewidth]{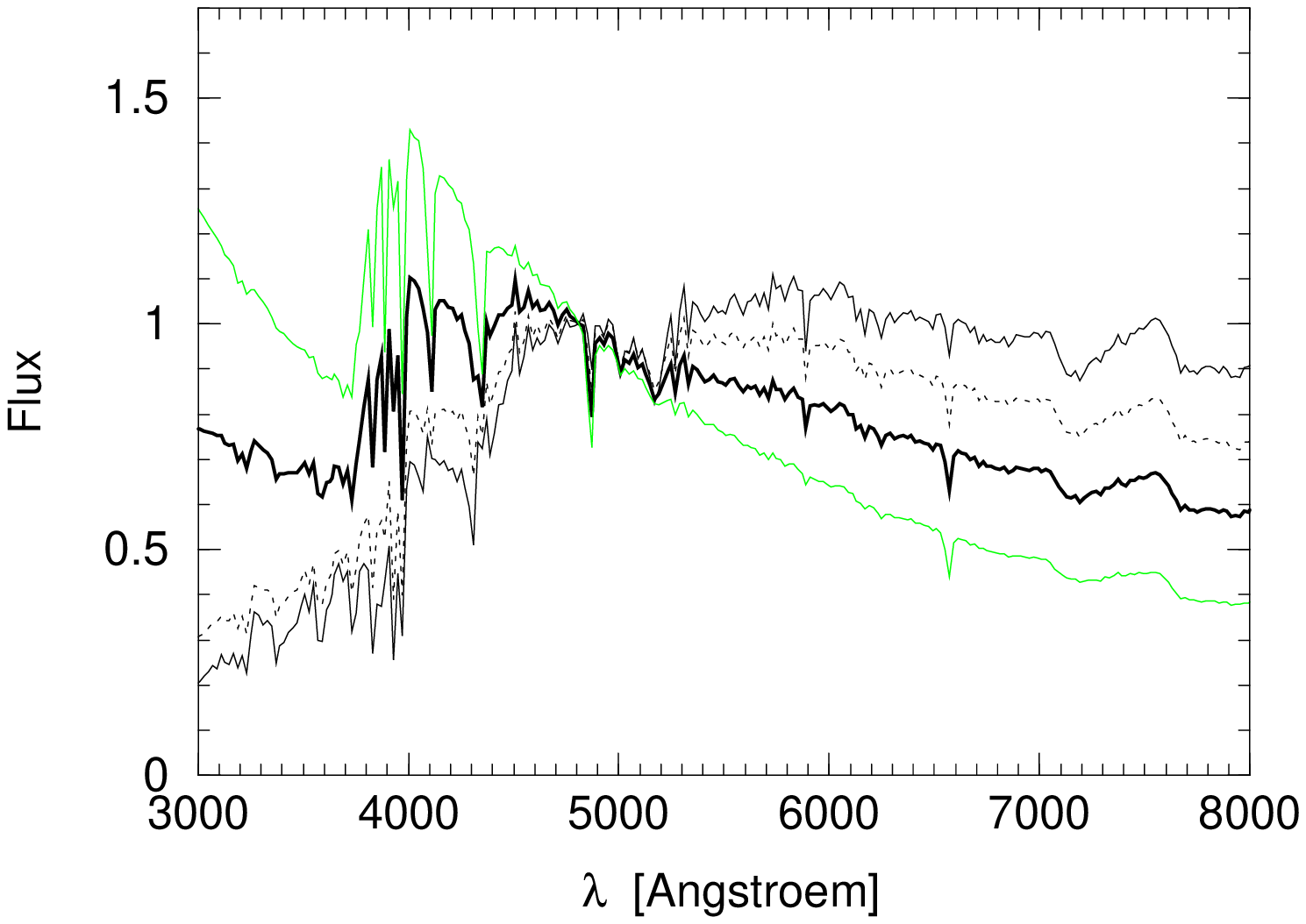}
   \caption{Contribution of the 3 phases of the 3-phase scenario to the total spectrum.
   {\itshape Left:\/} Absolute contribution (i.e., summation of the 3 subpopulation spectra gives the total
   spectrum). {\itshape Right:\/} Subpopulation spectra and total spectrum normalized at 4810\AA.}
   \label{abb.phasecontrib}
   \end{center}
\end{figure}

Figure \ref{abb.phasecontrib} illustrates the influence of the 3 different epochs of SF by showing, for the
SFH given above, the spectral contributions of stellar populations originating from each of these phases to the
total spectrum after 15 Gyrs:
The left panel shows, e.g., that phase 1, though its duration is almost three times longer than that of phase 2,
contributes only about half of the light of phase 2 to the total spectrum.
The right panel, on the other hand, shows that the first 14 Gyrs together have roughly the same influence on the
final shape of the spectrum as the most recent 1 Gyr of galaxy evolution.

\section{Some experimentation}
\begin{figure}[t]
   \begin{center}
   \includegraphics[width=0.32\linewidth]{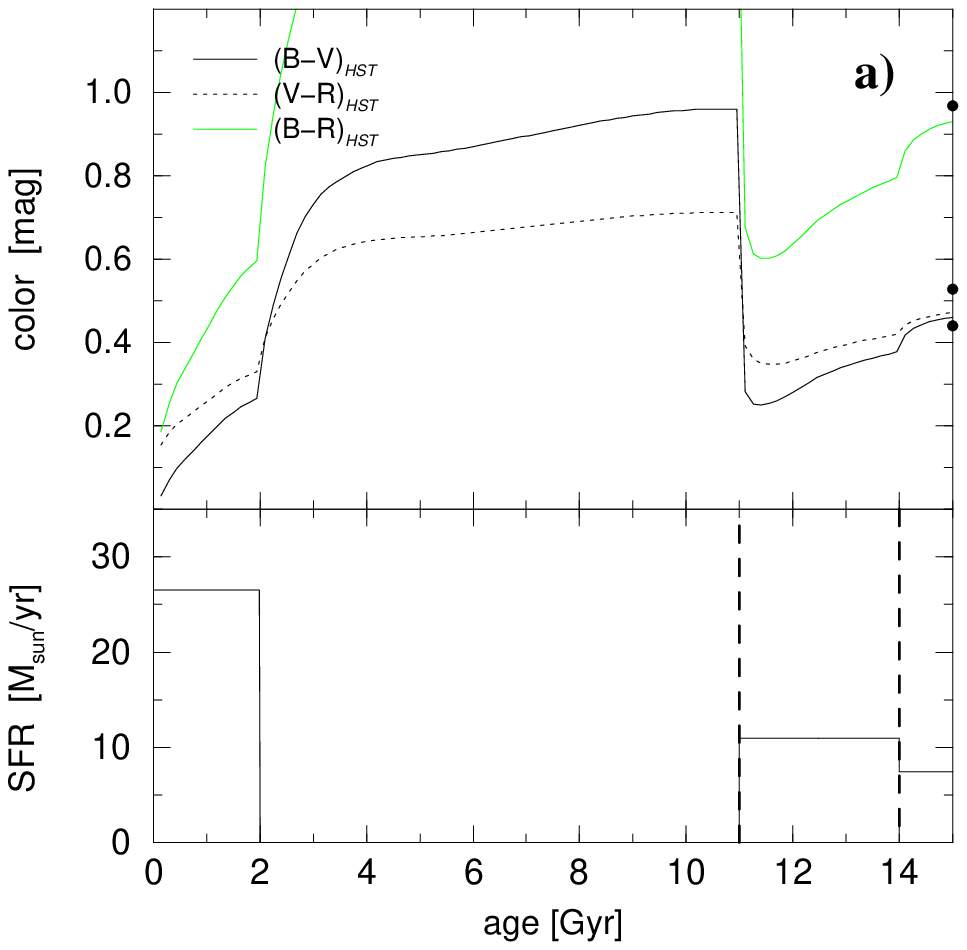} 
   \includegraphics[width=0.32\linewidth]{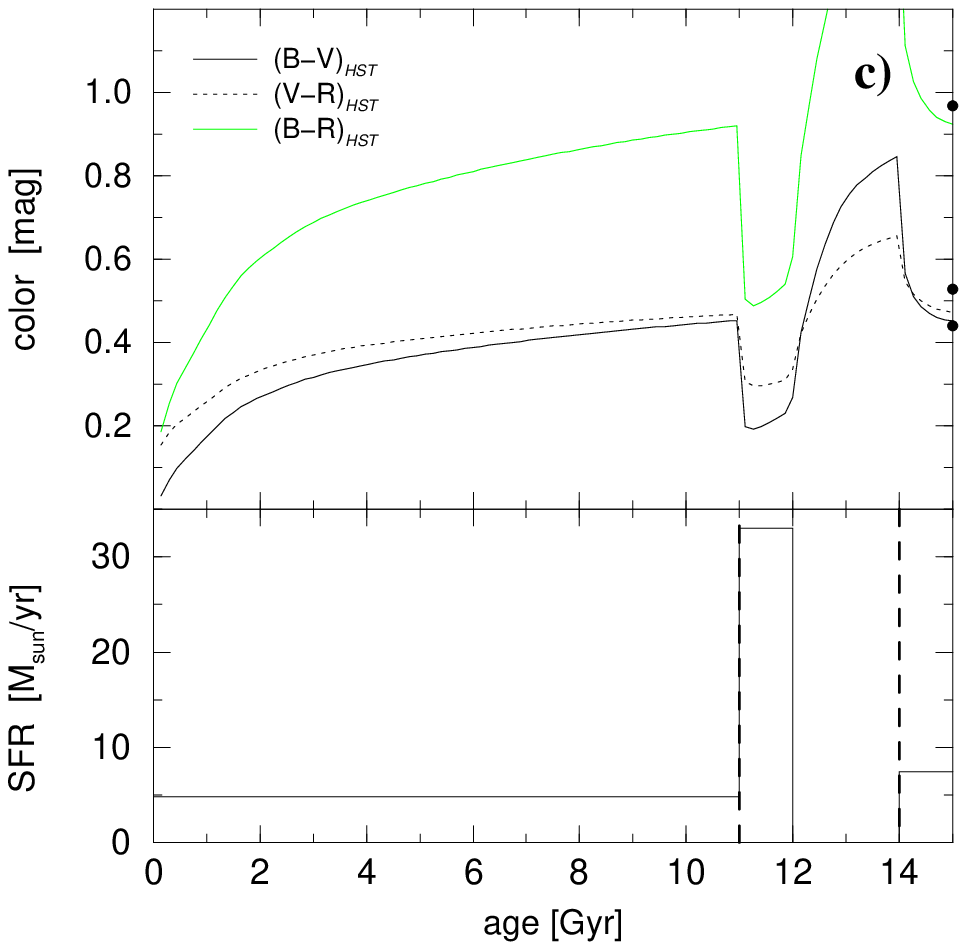} 
   \includegraphics[width=0.32\linewidth]{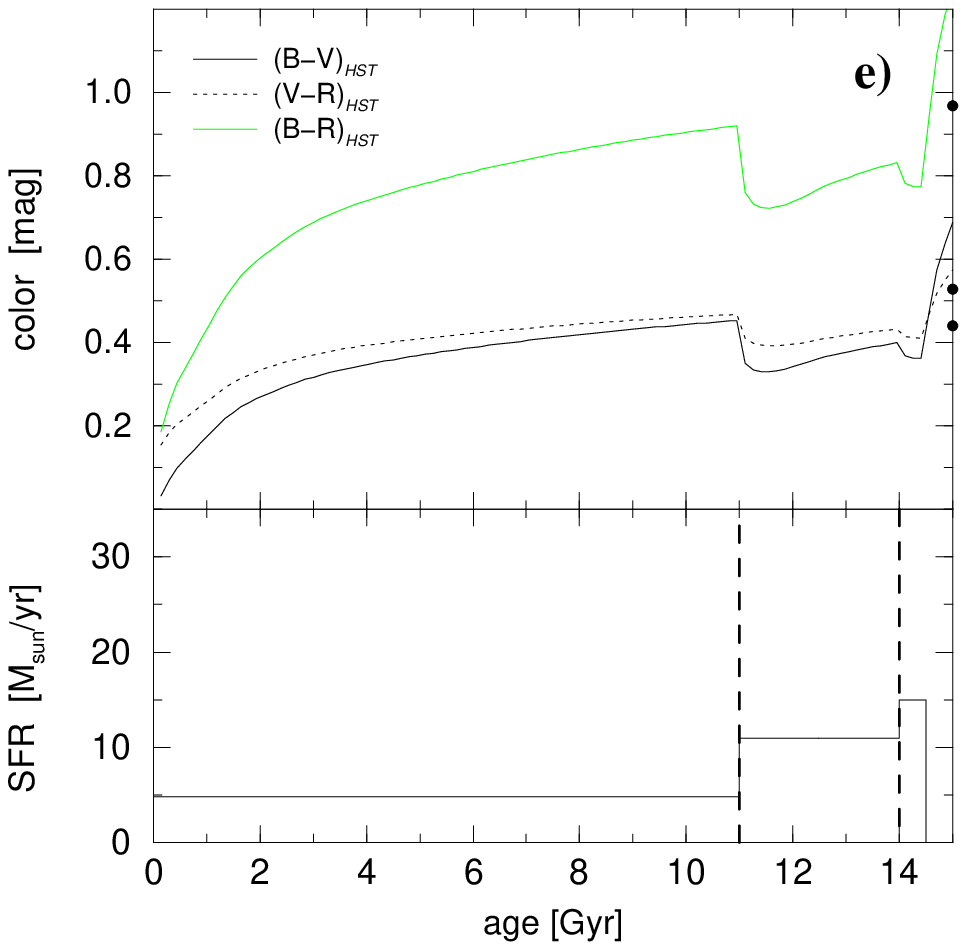} 
   \includegraphics[width=0.32\linewidth]{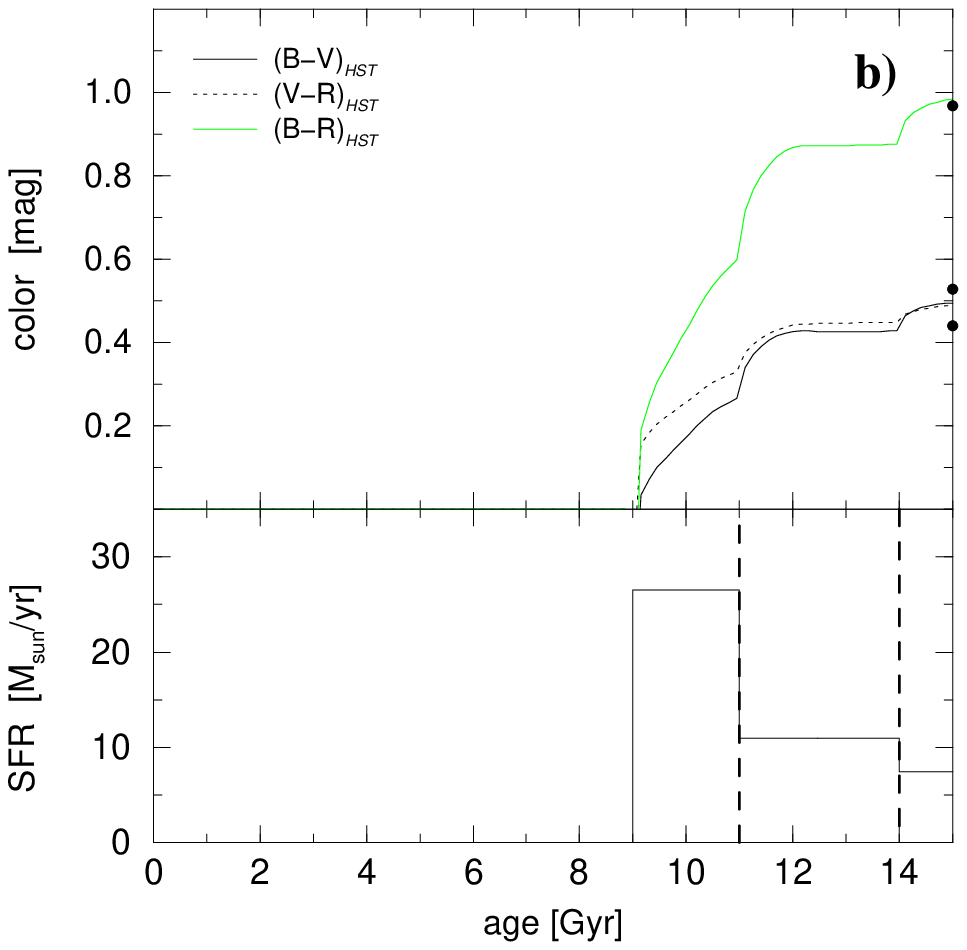} 
   \includegraphics[width=0.32\linewidth]{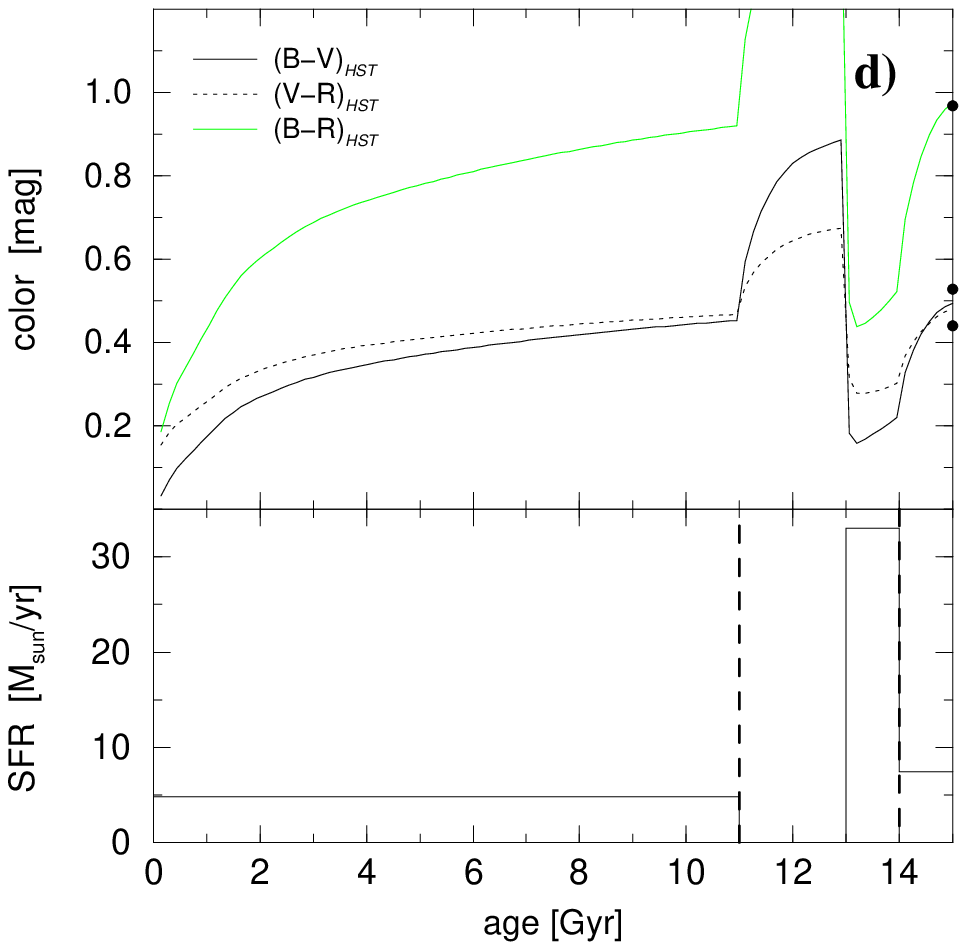} 
   \includegraphics[width=0.32\linewidth]{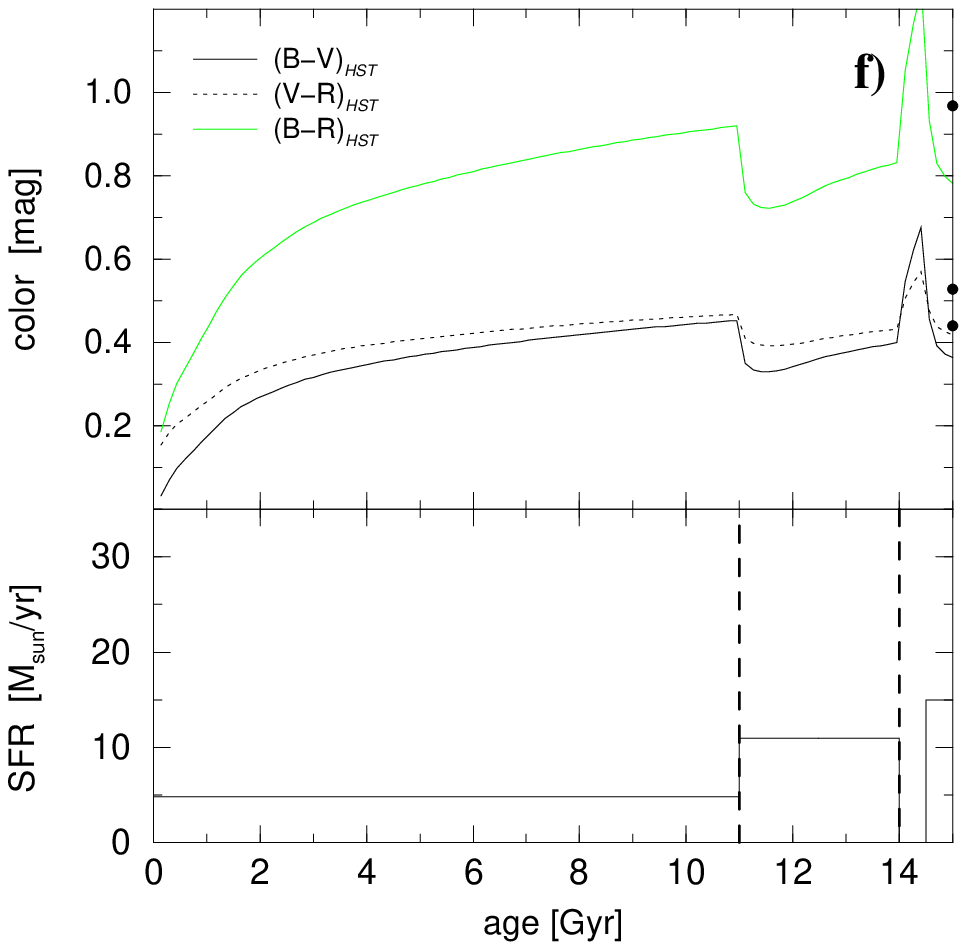} 
   \caption{Variations of the 3-phase scenario (cf. Fig. \ref{abb.lmc22}): Scenarios with systematic variations of
   the SFH within phase 1 ({\itshape a, b}), phase 2 ({\itshape c, d}), and phase 3 ({\itshape e, f});
   note that the relative distribution of the total amount of SF between the 3 phases remains unchanged.}
   \label{abb.lmc23bis28}
   \end{center}
\end{figure}
\begin{figure}[t]
   \begin{center}
   \includegraphics[width=0.32\linewidth]{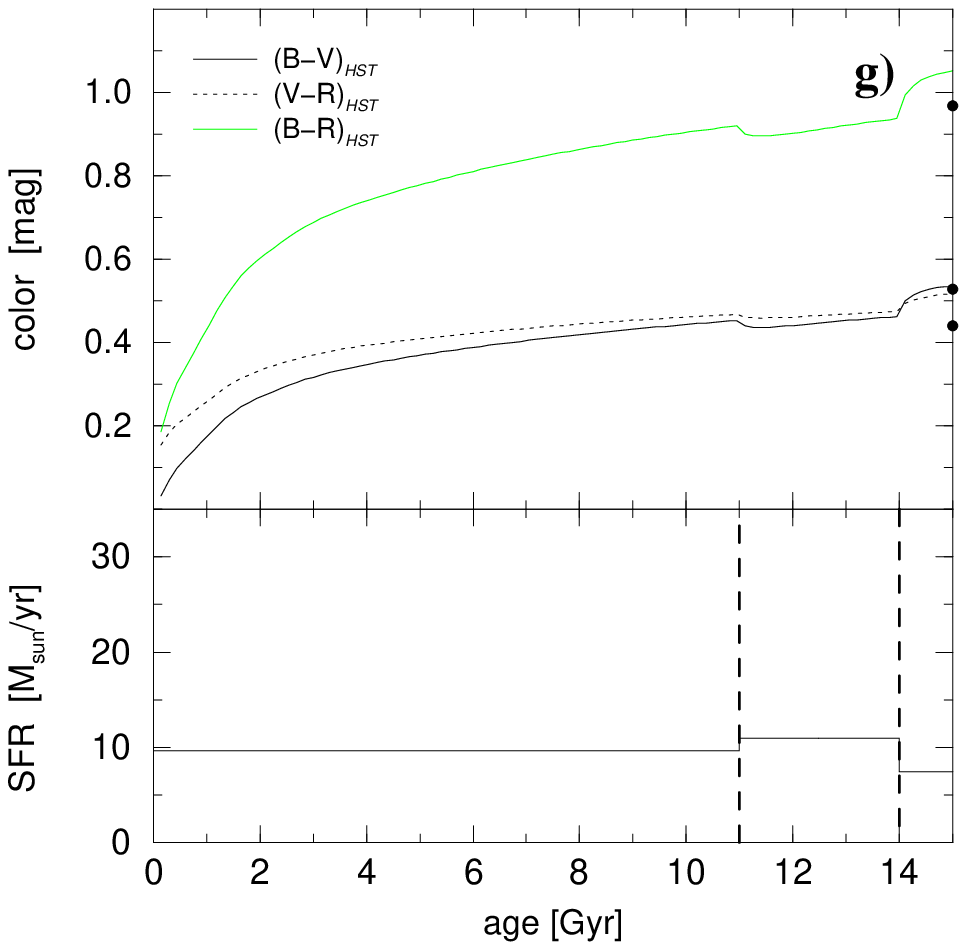} 
   \includegraphics[width=0.32\linewidth]{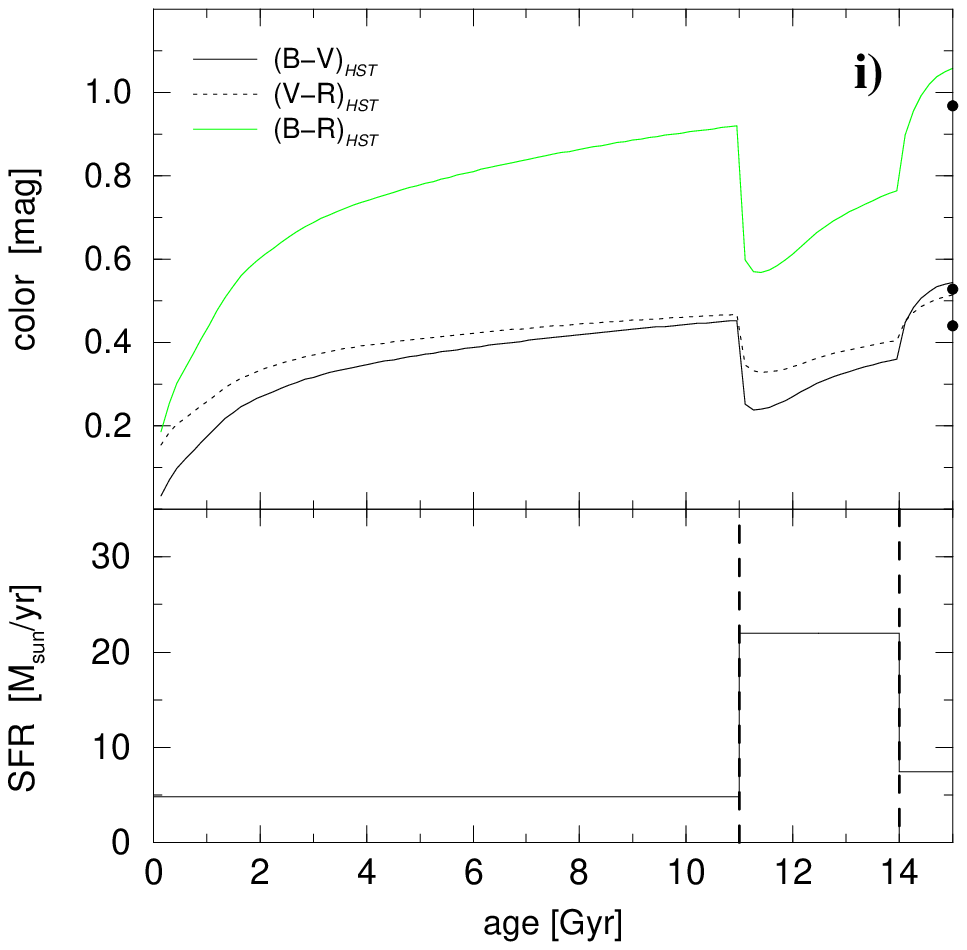} 
   \includegraphics[width=0.32\linewidth]{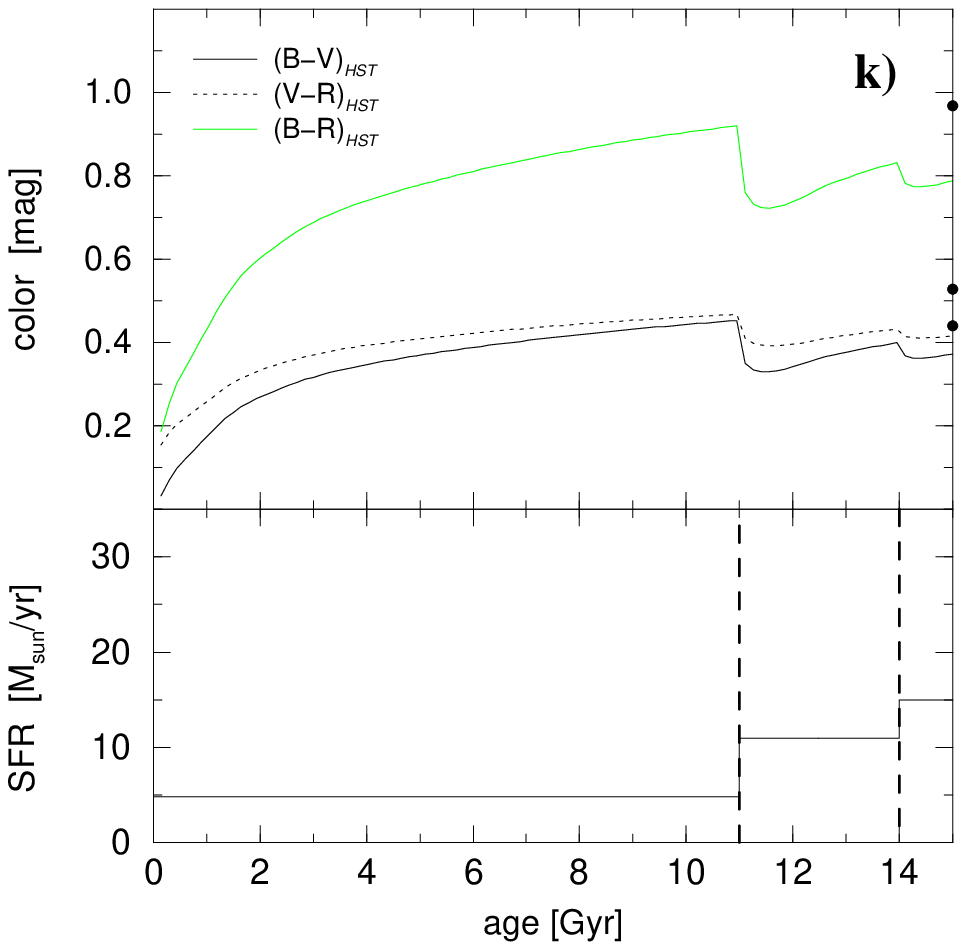} 
   \includegraphics[width=0.32\linewidth]{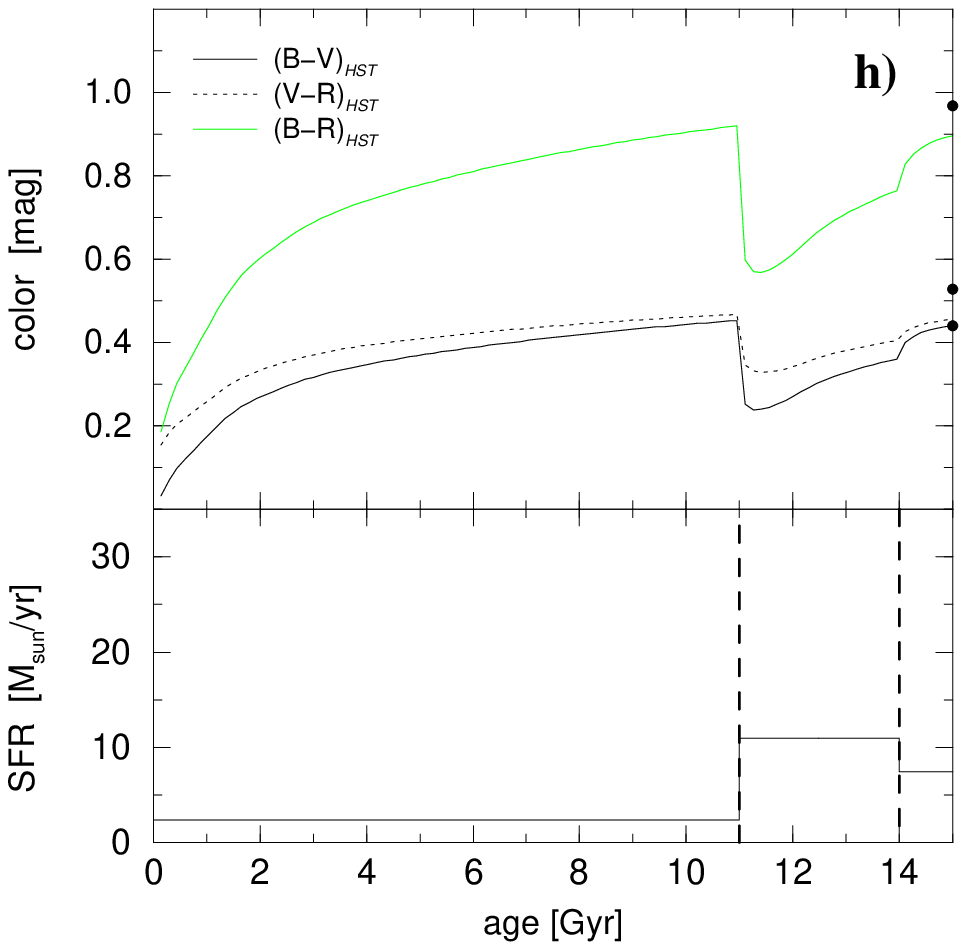} 
   \includegraphics[width=0.32\linewidth]{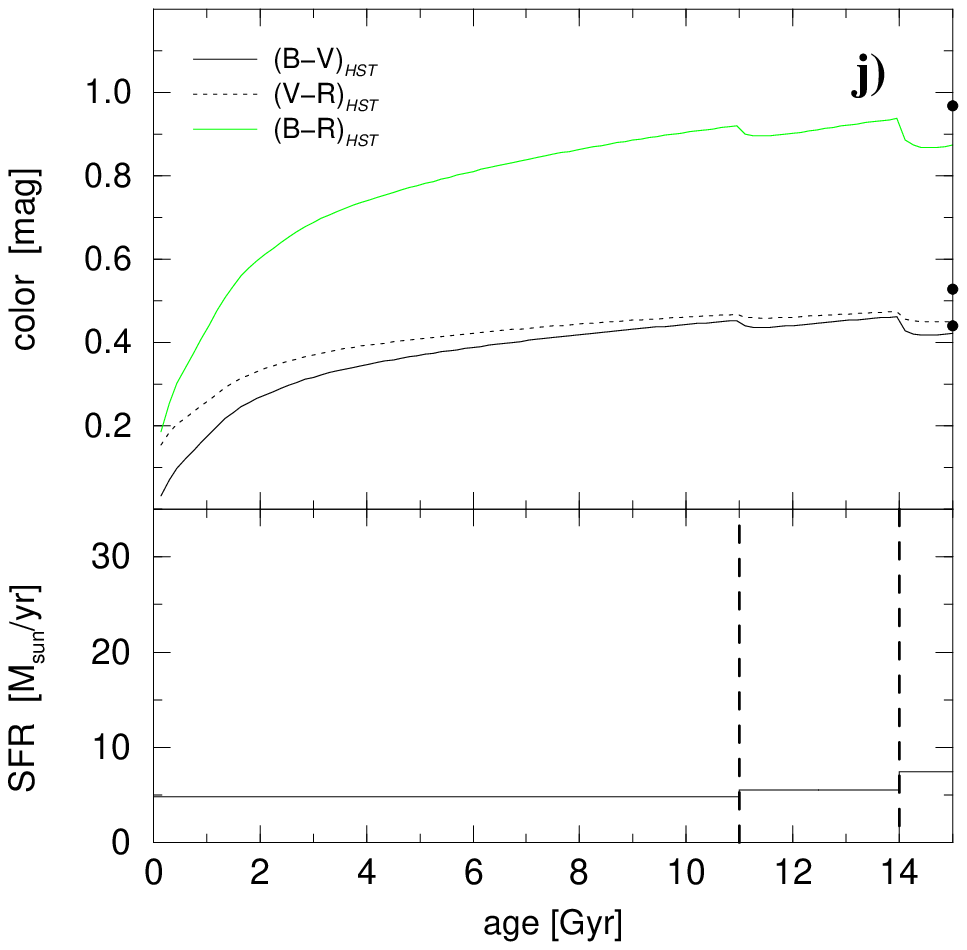} 
   \includegraphics[width=0.32\linewidth]{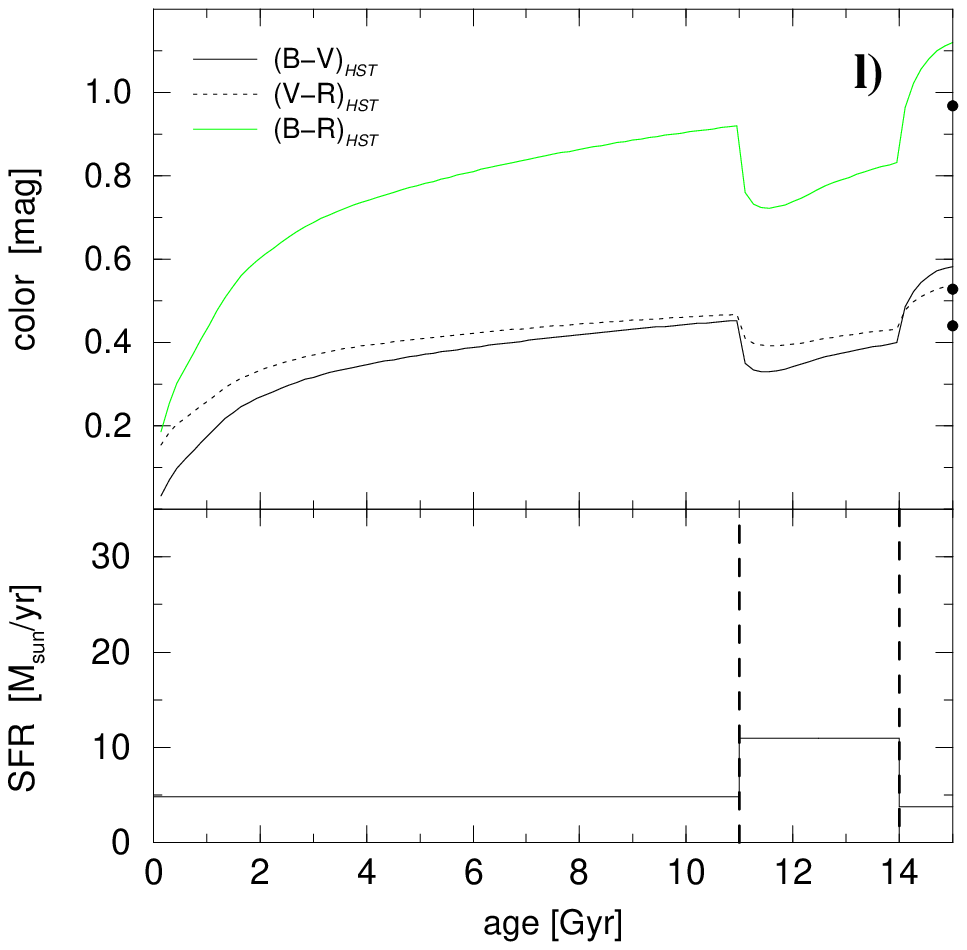} 
   \caption{Variations of the 3-phase scenario (cf. Fig. \ref{abb.lmc22}): Scenarios with twice and half the SF,
   respectively, within phase 1 ({\itshape g, h}), phase 2 ({\itshape i, j}), and phase 3 ({\itshape k, l});
   the relative distribution of the total amount of SF between the 3 phases is changed.}
   \label{abb.lmc22a1bis22c2}
   \end{center}
\end{figure}
We systematically vary the SFH in each of the 3 phases of our simple 3-phase scenario:\\

In the six scenarios shown in Figure \ref{abb.lmc23bis28}, the constant SF within the 3 phases is, for each phase,
replaced by a burst-like SF at the end and the beginning of the phase, respectively.
Because the relative distribution of SF \emph{between} the 3 phases remains unchanged, this kind of variations has
almost no effect on the final colors ($\Delta$color $<$ 0.05mag even in (B--R)$_{HST}$) if applied to phase 1 or 2;
remarkably, it is not even possible to decide if the galaxy is 15 or 6 Gyrs old in our example (cf. scenario
{\itshape a} and {\itshape b}).
The same kind of variation applied to phase 3 (i.e., to the most recent 1 Gyr of galaxy evolution), however,
results in a significant change of color ($\Delta$(B--R)$_{HST}$ $\approx$ 0.2mag).

In the scenarios shown in Figure \ref{abb.lmc22a1bis22c2}, the SF within the 3 phases remains constant, but,
compared to the original scenario, with half and twice the star formation rate (SFR), respectively. As before, the
variation is applied to each of the phases.
Since this kind of variation changes the relative distribution of SF between the 3 phases, the effect on the final
colors after 15 Gyrs is much larger than before, if applied to phases 1 and 2 ($\Delta$color $\approx$ 0.1mag in
(B--R)$_{HST}$). Variation of phase 3 (scenarios {\itshape k, l}) has a similar effect on colors
($\Delta$(B--R)$_{HST}$ $\approx$ 0.2mag) as the kind of variation applied in Fig. \ref{abb.lmc23bis28} (scenarios
{\itshape e, f}).\\

The experiments sketched above confirm the expectations from our previous work (cf. Section 3).
In phase 1 and 2, changes of the \emph{number of stars produced} within the phase have a stronger effect on colors
than changes of the \emph{mere distribution} of SF within the phase.
A similar behaviour for the effects of SFR changes in phases 1-3 is found in the spectral energy distribution (not
shown here).

\section{The CMD based SFH}
\begin{figure}[!ht]
   \begin{center}
   \includegraphics[width=0.9\linewidth]{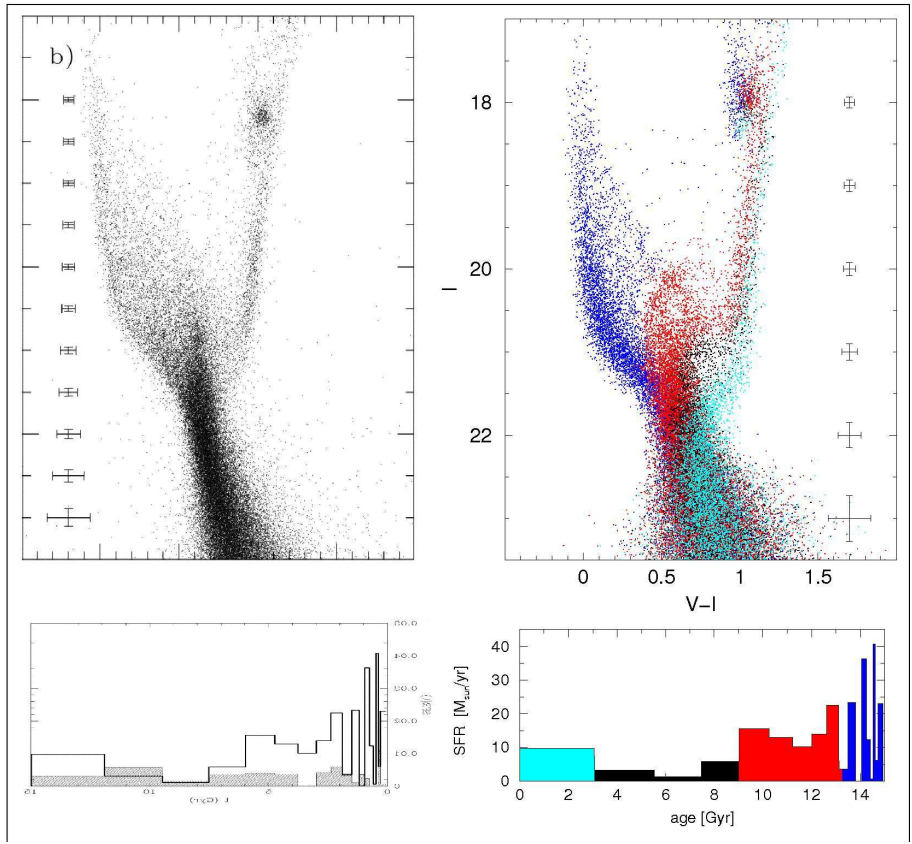}
   \caption{{\itshape Left:} CMD and SFH as presented by Smecker-Hane et al. (2002). {\itshape Right:} Model CMD at
   a simulated galaxy age of 15 Gyr, using Smecker-Hane et al.'s SFH. Stellar populations originating from 4
   different phases of star formation are coded in different colors (cf. the electronic version of this paper).}
   \label{abb.CMDs}
   \end{center}
\end{figure}
The resolved stellar population of the LMC bar field was observed with \emph{HST} WFPC2 in 1997 (PI:
Smecker-Hane).\\

Smecker-Hane et al. (2002) presented a SFH derived from their analysis of the high-quality CMD obtained from these
observations; Figure \ref{abb.CMDs} (left) shows this CMD together with the published SFH\footnote{Note that the
time axis is inverted to match the standard used in this paper (evolution of the galaxy from left to right).}. The
right panel of the same figure shows a model CMD at a simulated galaxy age of 15 Gyr, using the SFH from
Smecker-Hane et al. (2002), with 4 color-coded epochs of star formation (cf. the electronic version of this
contribution on astro-ph).

Note that the CMD is computed using a relatively simple approach (cf. Lilly 2003):
The number of stars at each point on the isochrones is determined by the IMF and the relative weight of the
isochrone; the stars are spread around their theoretical position by applying observational errors in I and
V-I, small for bright stars and larger for fainter stars.
For any given SFH (and chemical enrichment history) our code is able to calculate the \emph{time evolution} of the
distribution of stars in the HR diagram and any desired CMD.
However, we do not interpolate between isochrones; therefore, we had to increase the assumed observational errors
in order to reduce the ``gaps'' between isochrones on the CMD (cp. observed with modell errors as shown in Fig.
\ref{abb.CMDs}). Also, features like binary stars are not regarded.

Hence, our model CMDs are not intended to be directly compared with observations but, so far, for principle
investigations only (model-model comparisions).\\

Figures \ref{abb.smeckerhane} and \ref{abb.smeckerhaneCMD}, left panels, show the spectrophotometric evolution
and the model CMD at 15 Gyr galaxy age resulting from Smecker-Hane et al.'s SFH.
The middle panels of the same figures show the same for a scenario using a ``smoothed'' version of this SFH.
``Smoothed'' means that within each of the four epochs of SF (color coded in Fig. \ref{abb.CMDs}), the SFR is
put to a constant value, conserving the number of stars produced in this epoch (i.e., without changing the
relative amounts of SFR \emph{between} the 4 epochs). Note, however, that the latter is not the case when
compared to the 3-phase SFH (right panels in Figs. \ref{abb.smeckerhane} and \ref{abb.smeckerhaneCMD}).

In the next section, both scenarios will be confronted against each other and against the simple 3-phase scenario.

\section{CMDs and integrated light: Comparison and Conclusions}
\begin{figure}[t]
   \begin{center}
   \includegraphics[width=0.32\linewidth]{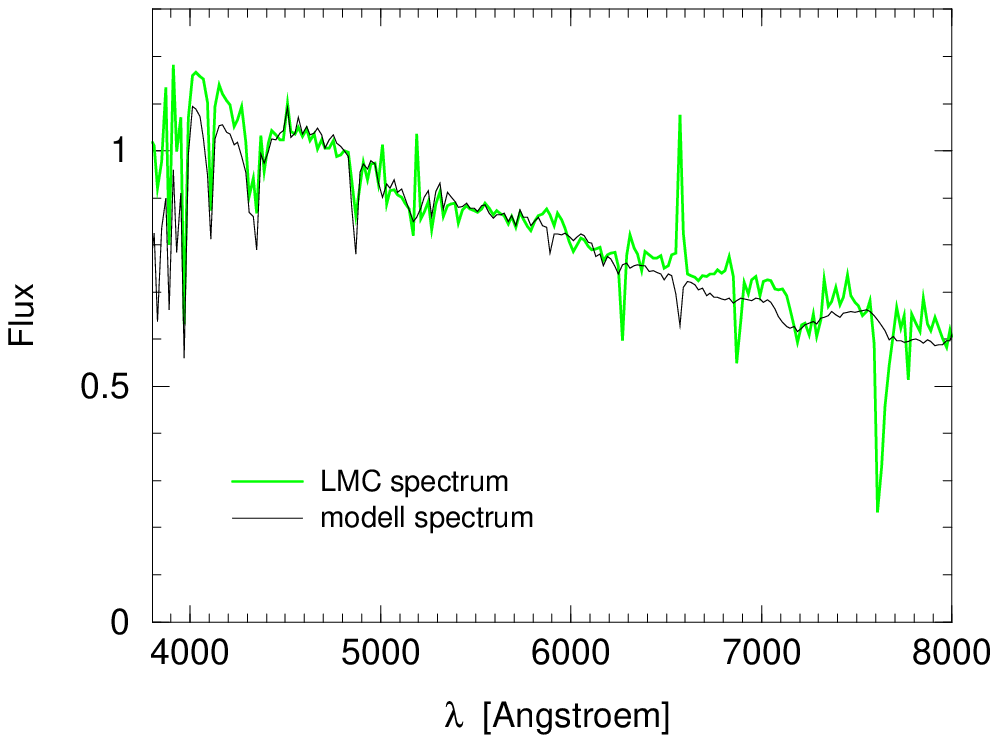}
   \includegraphics[width=0.32\linewidth]{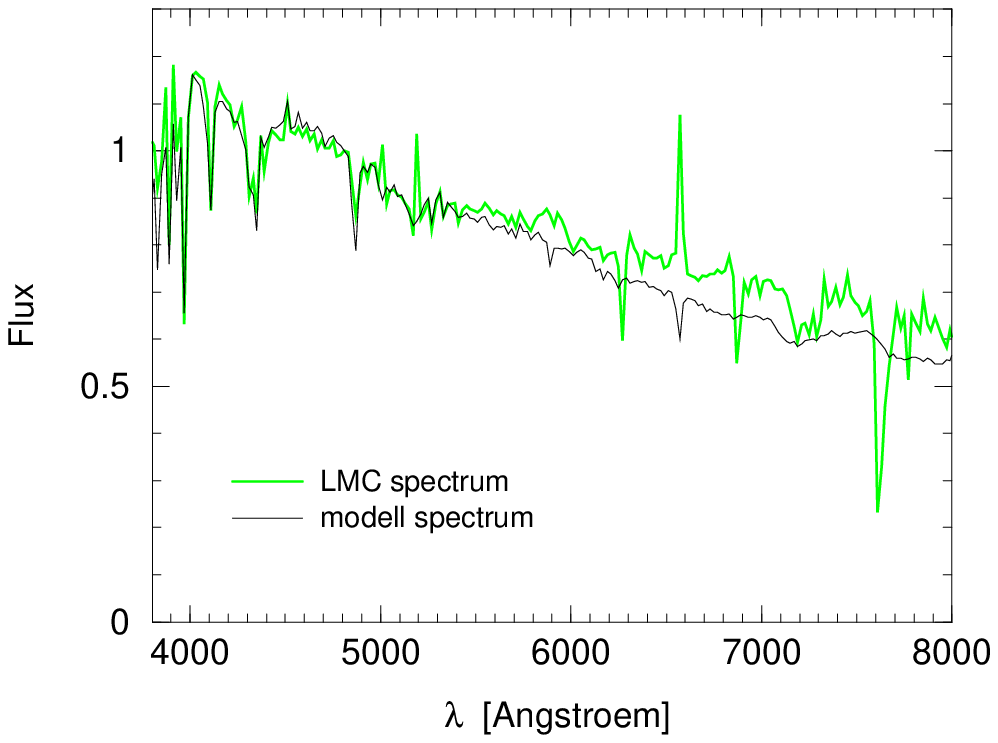}
   \includegraphics[width=0.32\linewidth]{x.spec_lmc22.eps}
   \includegraphics[width=0.32\linewidth]{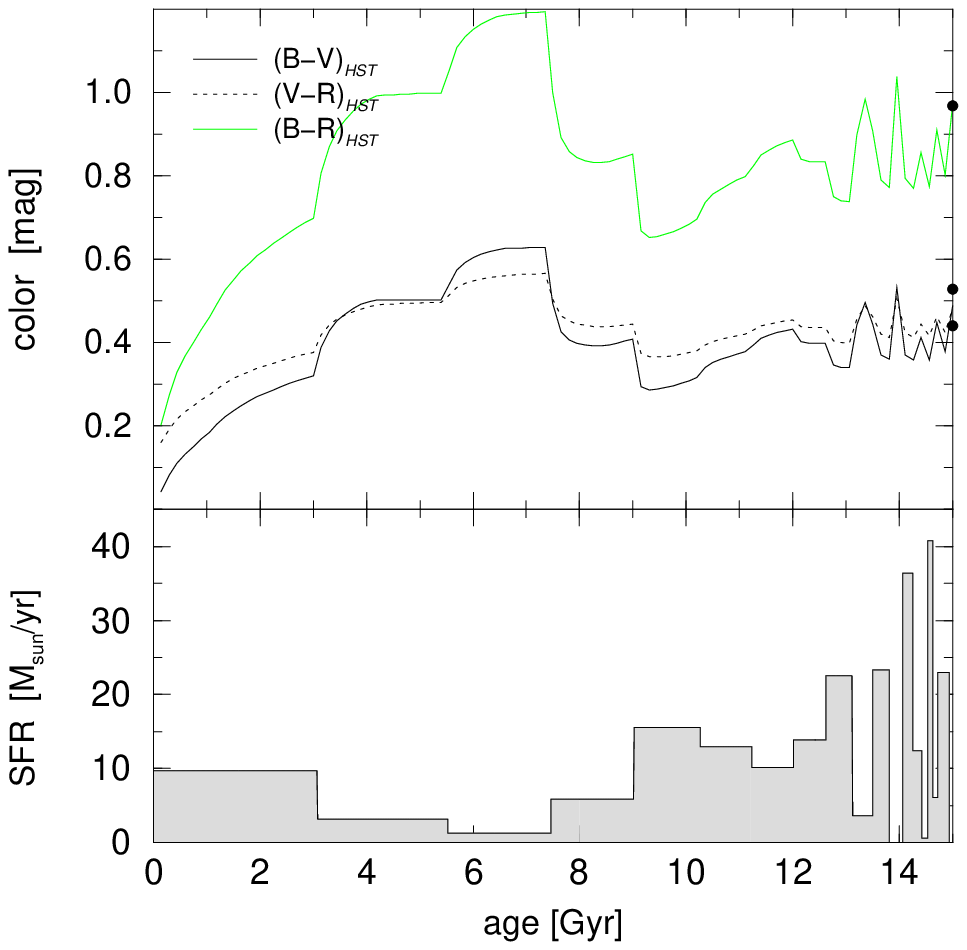}
   \includegraphics[width=0.32\linewidth]{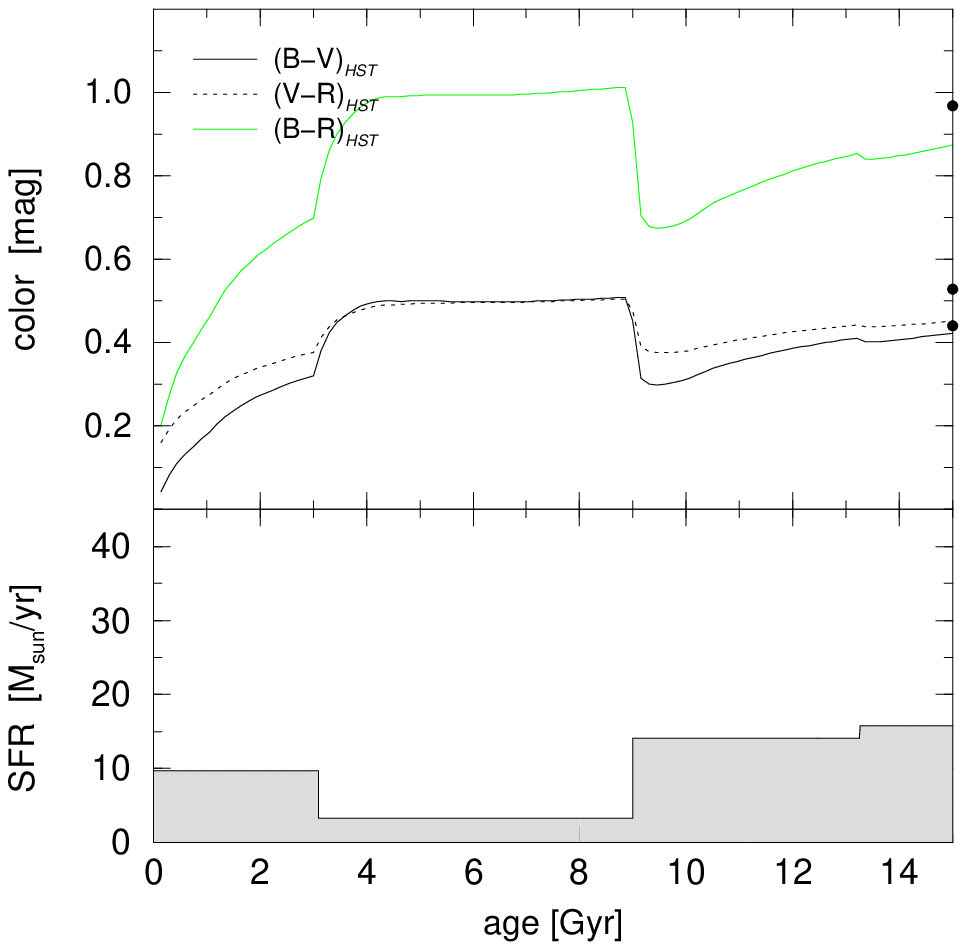}
   \includegraphics[width=0.32\linewidth]{x.Dipl22.eps}
   \caption{Confrontation of scenarios using Smecker-Hane at al.'s (2002) original SFH (left panels), a
   ``smoothed'' Smecker-Hane SFH (central panels), and using a simple 3-phase SFH (right panels); see text.\newline
   {\itshape Top:\/} Model spectra of the scenarios after 15 Gyr (black) against observed spectrum (grey).
   {\itshape Bottom:\/} Photometric evolution of the scenarios in terms of (B-V)$_{HST}$, (V-R)$_{HST}$, and
   (B-R)$_{HST}$ with the corresponding SFHs; the observed colors (obtained from the observed spectrum) are marked
   with black dots at 15 Gyr.}
   \label{abb.smeckerhane}
   \end{center}
\end{figure}
\begin{figure}[t]
   \begin{center}
   \includegraphics[width=0.32\linewidth]{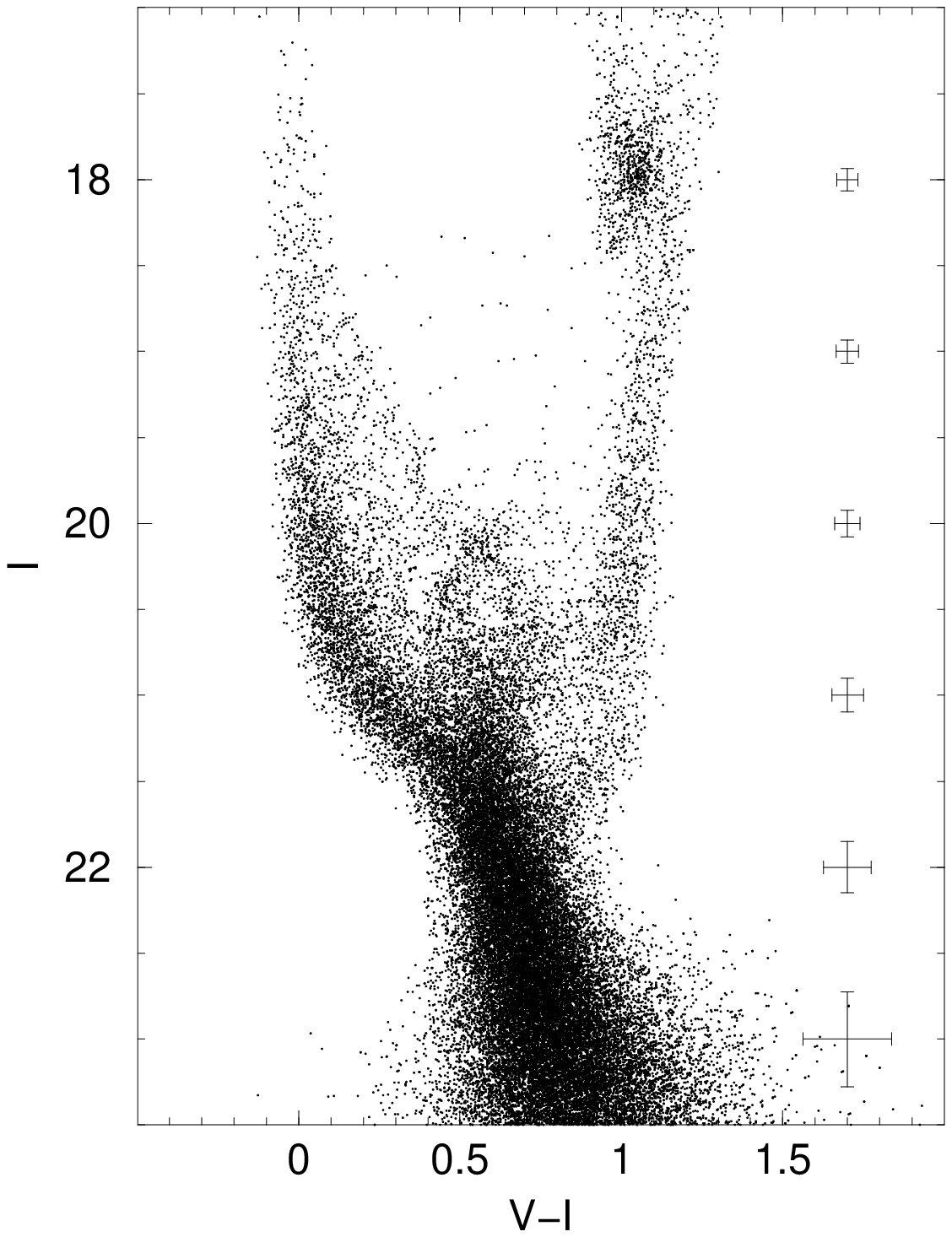}
   \includegraphics[width=0.32\linewidth]{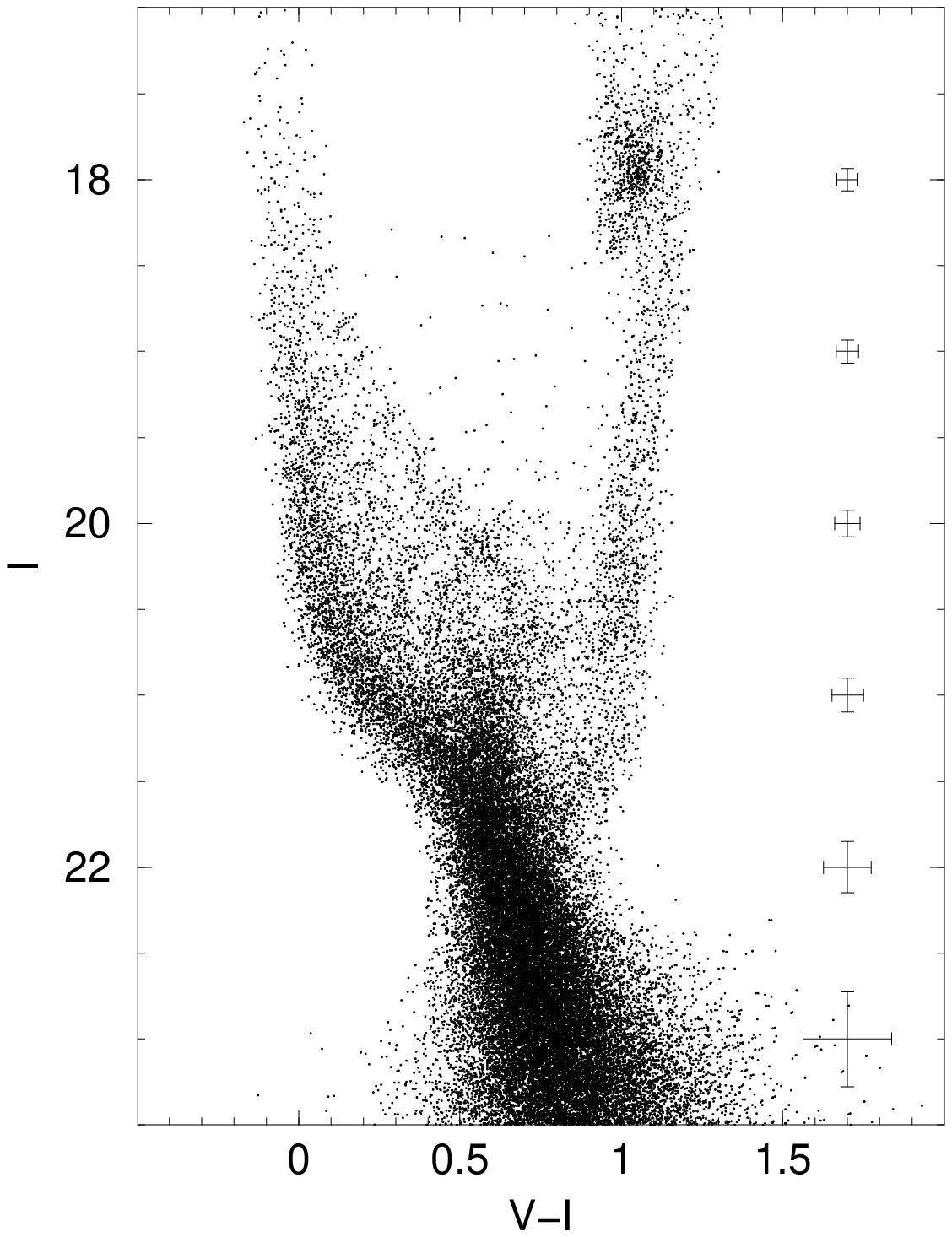}
   \includegraphics[width=0.32\linewidth]{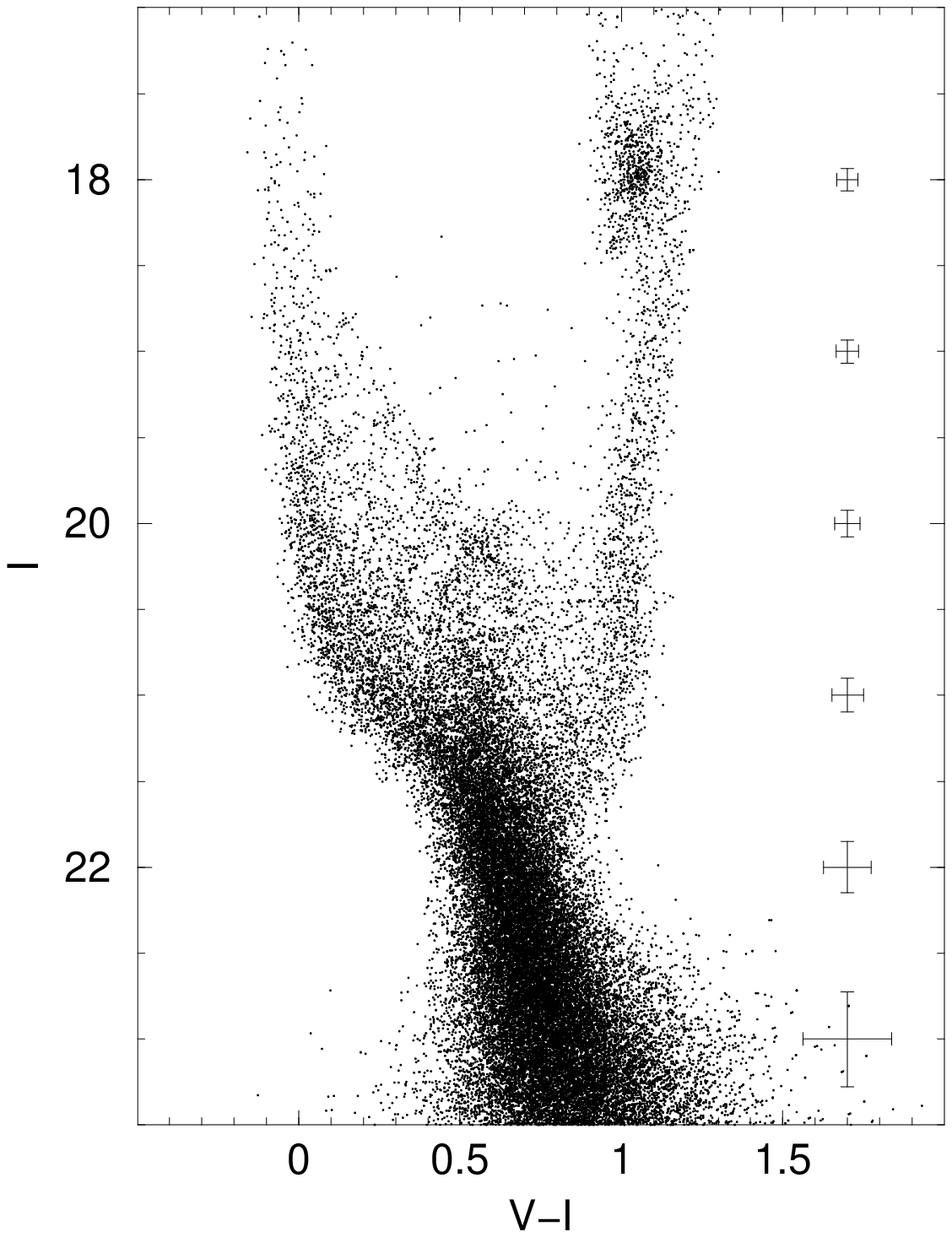}
   \includegraphics[width=0.32\linewidth]{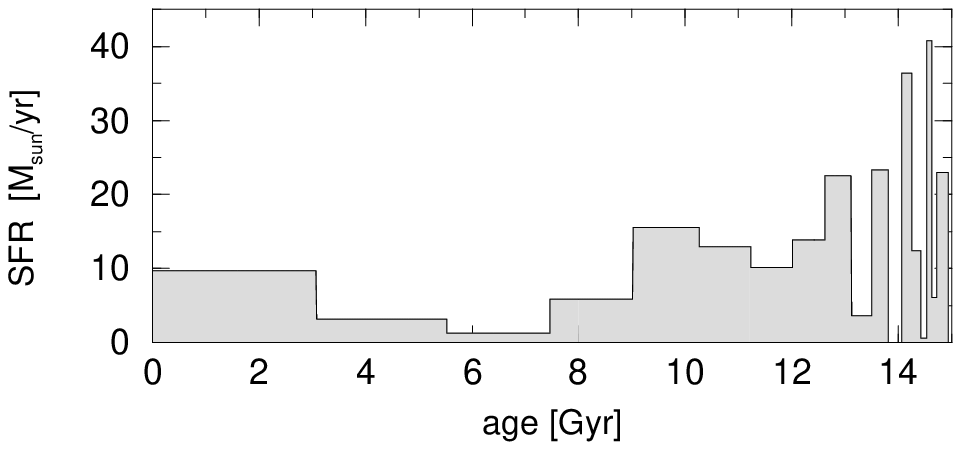}
   \includegraphics[width=0.32\linewidth]{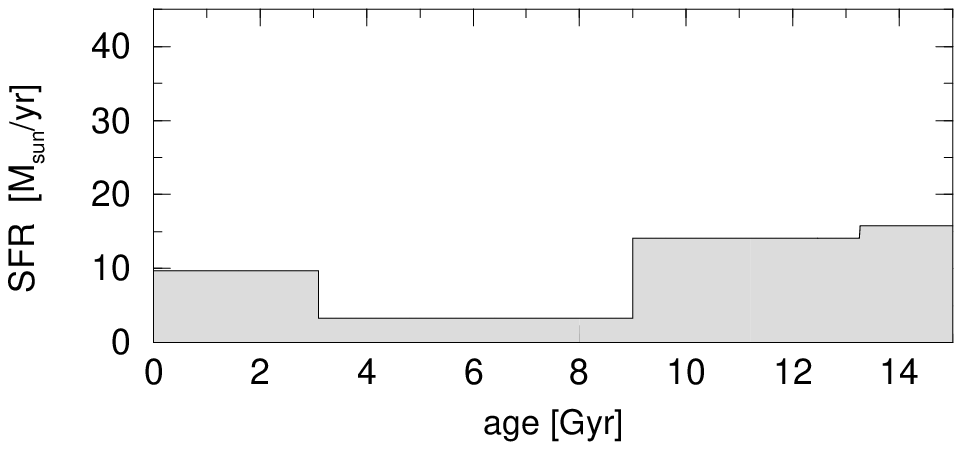}
   \includegraphics[width=0.32\linewidth]{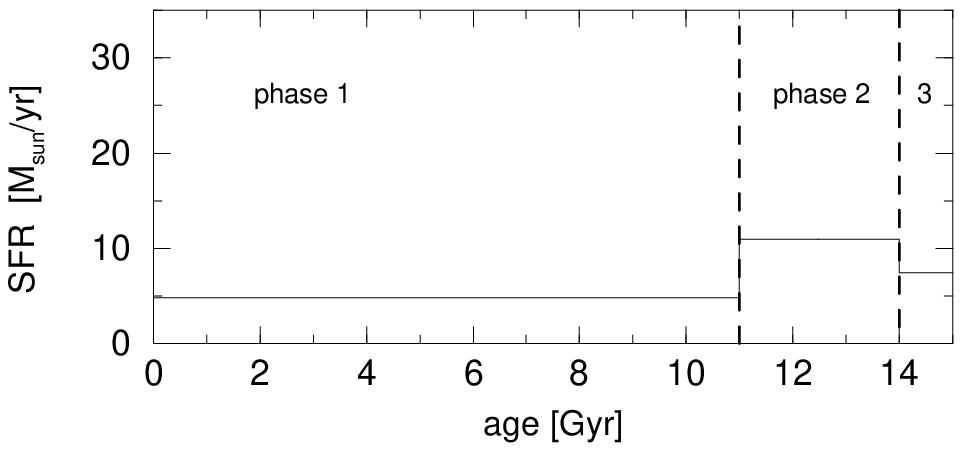}
   \caption{Confrontation of scenarios using Smecker-Hane et al.'s (2002) original SFH (left panels), a
   ``smoothed'' Smecker-Hane SFH (central panels), and using a simple 3-phase SFH (right panels); see text.\newline
   {\itshape Top:\/} Model CMDs at a simulated galaxy age of 15 Gyr. {\itshape Bottom:\/} Corresponding SFHs.}
   \label{abb.smeckerhaneCMD}
   \end{center}
\end{figure}
Figures \ref{abb.smeckerhane} and \ref{abb.smeckerhaneCMD} confront the spectrophotometric evolution and the CMDs,
respectively, of three scenarios using Smecker-Hane et al.'s (2002) SFH (left panels), using the ``smoothed''
Smecker-Hane SFH (central panels), and using the ``simple 3-phase'' SFH (right panels).\\

In terms of colors and spectra, the scenario using Smecker-Hane et al.'s SFH is, at a galaxy age of 15 Gyr,
practically identical to the 3-phase scenario; both differ slightly ($\Delta$(B--R)$_{HST}$ $<$ 0.1mag) from the
``smoothed'' Smecker-Hane scenario.

This shows that two scenarios with very different SFHs can result in very similar, observationally
indistinguishable integrated-light properties; on the other hand, scenarios with very similar ``global'' SFH (as
e.g. the ``original'' and ``smoothed'' Smecker-Hane et al. SFH) can differ in their final colors.
The results show not only the ambiguity of SFHs obtained from integrated light but also emphasize again the
importance of the \emph{most recent} (i.e., lookback time $\leq$ 1 Gyr) SFR for the observed colors.\\

In terms of CMDs, the ``original'' and ``smoothed'' Smecker-Hane et al. scenarios seems to be very similar; the
CMD computed using the 3-phase scenario differs from both.
If this result can be validated by a solid numerical comparision between the CMDs (not done yet), this means that
the \emph{global} distribution of SF is crucial for the appearance of CMDs - as opposed to integrated light.

However, note that not even CMDs are free of ambiguity. The ambiguity in the SFH derived from a CMD increases with
increasing lookback time, and SFHs like that presented by Smecker-Hane et al. (2002) are most likely not as exact
as their complicated shape suggests; this was already shown by Lilly (2003) and most recently confirmed by the
``double-blind Cozumel experiment'' (Holtzman 2005, these proceedings).\\
%

\acknowledgements             
TL gratefully acknowledges partial travel support from the organizers; 
his work is partially funded by DFG grant Fr 916/11-1-2-3.\\



\end{document}